\documentclass[aps,prb,twocolumn,superscriptaddress,amsmath,amssymb]{revtex4-1}
\usepackage{graphicx}
\usepackage{epsf}
\usepackage{bm}

\def\ls{\lesssim}

\begin{document}
\title{
Connecting thermodynamic and 
dynamical anomalies of water-like liquid-liquid phase transition in the
 Fermi--Jagla model
}

\author{Saki Higuchi}
\thanks{Equally contributed to this work.}
\author{Daiki Kato}
\thanks{Equally contributed to this work.}

\author{Daisuke Awaji}
\affiliation{
Graduate School of Science and Technology, Niigata University, Niigata 950-2181, Japan
}

\author{Kang Kim}
\email{kk@cheng.es.osaka-u.ac.jp}
\affiliation{
Division of Chemical Engineering,
Department of Materials Engineering Science, Graduate
School of Engineering Science, Osaka University, Toyonaka, Osaka 560-8531, Japan
}

\date{\today}

\begin{abstract}
We present a study using molecular dynamics simulations based on the Fermi--Jagla
     potential model, which is the continuous version of the mono-atomic core-softened
     Jagla model [J. Y. Abraham, S. V. Buldyrev, and
     N. Giovambattista, J. Phys. Chem. B, \textbf{115}, 14229 (2011)].
This model shows the water-like liquid-liquid phase transition between
     high-density and low-density liquids at the liquid-liquid critical point.
In particular, the slope of the coexistence line becomes weakly
     negative, which is expected to represent one of anomalies of liquid
 polyamorphism.
In this study, we examined the density, dynamic, and thermodynamic
     anomalies in the vicinity of the liquid-liquid critical point.
The boundaries of density, self-diffusion, shear viscosity, and
     excess entropy anomalies were characterized.
Furthermore, these anomalies are connected according to the Rosenfeld's
     scaling relationship between the excess entropy and the transport
     coefficients such as diffusion and viscosity.
The results demonstrate the hierarchical and nested structures
     regarding the thermodynamic and dynamic anomalies of the Fermi--Jagla model.
\end{abstract}

\maketitle

\section{Introduction}

Liquid polyamorphism resulting from liquid-liquid phase transitions
(LLPTs) of a one-component liquid system is
an importance scenario associated with
various anomalous properties of liquid
water.~\cite{Poole:1992ka, Angell:1995dp, Stanley:1998hf, 
Debenedetti:2003gd, Debenedetti:2003gn, Stanley:2013gw, Nayar:2013dl,
Gallo:2016fd}
This scenario suggests that the first-order phase transition
distinguishes between two liquid states with different densities,
namely, a high-density liquid (HDL) and a low-density liquid (LDL).
The polymorphs eventually vanish at the liquid-liquid critical point (LLCP),
which is known as a second critical point that is different from the
liquid-gas critical point.
In addition, the anomalies in liquid states and the possible scenario of the
LLPT have been widely studied in other substances such as silicon,~\cite{Sastry:2003dr,
Molinero:2006hk, Ganesh:2009it,
Beye:2010gq, Vasisht:2011ch}, silica~\cite{Poole:1997it, SaikaVoivod:2000ff,
SaikaVoivod:2004hq, Lascaris:2014jq, Chen:2017ek}
germanium~\cite{Bhat:2007er, Hujo:2011ct, Li:2016bq},
germania~\cite{Jabes:2010ht}, and 
phosphorous.~\cite{Katayama:2000cz, Morishita:2001gg}

To clarify the universal mechanism of the LLPT in liquid water,
intensive simulation studies have been carried out using realistic water
models, and the results have generated much
controversy.~\cite{Molinero:2009kc, Liu:2009bt, Sciortino:2011gf, Limmer:2011fu,
Liu:2012bt, Kesselring:2013by, Sumi:2013fy, Limmer:2013iq,
Overduin:2013cu, Palmer:2014jb, Russo:2014fl,
Yagasaki:2014ew, Smallenburg:2015hw, Overduin:2015tw,Singh:2016bu, Biddle:2017bb}
Other simulation studies on LLPTs and water-like anomalies utilize the
short-ranged and isotropic pair potential, which is contrary to the
above-mentioned model of rigid-body water molecules.
The family of such potential models is known as ``core-softened
potentials,'' which originate from the studies by
Stell and Hemmer.~\cite{Hemmer:1970vr, Stell:1972ja} and those by
Jagla~\cite{Jagla:1999dk, Jagla:2001kd, Jagla:2001kn}
In comprehensive and extensive numerical studies,
the core-softened potential models serve as good reference models to reveal the
underlying mechanism of LLPTs and water-like
anomalies.~\cite{SadrLahijany:1998vr, Franzese:2001fy, Xu:2005ha, Xu:2006fp, Xu:2006ff, 
Errington:2006dl, Sharma:2006fy, Gibson:2006cj, Yan:2006un,
Stanley:2007gf, Yan:2008un, Stanley:2008bn, Fomin:2008it,
deOliveira:2008ga, Buldyrev:2009dk, Gribova:2009tp, deOliveira:2009kx,
Xu:2010cu, BarrosdeOliveira:2010ib, Salcedo:2011di, Barraz:2011ht, Fomin:2011fz, Fomin:2011gb,
Fomin:2013ef, Fomin:2013fa, Luo:2014jw, Luo:2015ku,
Ricci:2017ik, Pinheiro:2017fv}
The core-softened potential generally involves two potential minima with
different length scales, which mimics the effective
potential between two water pentamers.~\cite{Scala:2000im,
Buldyrev:2002tt, Stanley:2007gf}
Furthermore, the two-scale Jagla potential is proposed, which
consists of a repulsive hard-core potential and an attractive potential comprising two linear
ramps.
It has been demonstrated that this potential exhibits the LLPT and the associated
water-like anomalies in density, diffusion, structures, and
thermodynamic functions.~\cite{Xu:2005ha, Xu:2006fp, Xu:2006ff}

The hard-core part in the Jagla potential is described by a step
function, which requires an event-driven method for molecular
dynamics simulations.
As an alternative,
the Fermi--Jagla (FJ) model has recently been proposed for simulations based on
finite discrete time-step molecular dynamics.~\cite{Abraham:2011bx,
Gordon:2014fo, Sun:2015jj, Sun:2017wo, Sun:2018ff}
The hard-core part of the model is described by a soft-core potential and
the two-ramp part is replaced with two Fermi distribution functions.
It has been demonstrated that the FJ model also exhibits the
water-like anomalies such as anomalous trends in thermodynamic and dynamic behaviors.
In addition, the LLPT and LLCP can be determined from a
pressure-temperature phase diagram.
Contrary to the original Jagla potential, the slope of the coexistence line between
the HDL and LDL phases becomes slightly negative,
which is expected to be one important characteristic of the liquid-water anomaly.
The similar negative slope associated with the LLPT has been observed in
simulations using
ST2 model~\cite{Xu:2005ha, Xu:2006ff, Xu:2006fp} and Jagla-like potentials.~\cite{Gibson:2006cj, Luo:2014jw, Luo:2015ku}

In this paper, we report the molecular dynamics simulation results
obtained with the FJ model, which are along the previous study~\cite{Abraham:2011bx}.
We not only reproduce previously reported results including LLPT and
LLCP, but also provide further numerical results.
First, the dynamic anomaly of the FJ model was comprehensively
characterized by the shear viscosity and the self-diffusion coefficient.
The quantification of two transport coefficients allowed us to
examine 
the breakdown of the Stokes--Einstein (SE) relationship, as in the
literature of glass-forming liquids.
We revealed that the degree of the SE breakdown depends on the density,
\textit{i.e.}, HDL or LDL.
This implies that the SE breakdown is related to the concept of fragility, which
characterizes a degree of Arrhenius behavior in the temperature
dependence of the glassy dynamics.
Next, the thermodynamic anomaly was identified by the density
dependence of excess entropy that was calculated from thermodynamic
integration calculations, in contrast to the two-body excess entropy approximation.
The connection between anomalies in thermodynamics and transport
coefficients was demonstrated using Rosenfeld's excess entropy scaling
relationship.~\cite{Rosenfeld:1977tu, Rosenfeld:1999ii}
This scaling analysis was useful to predict 
the nested dome-like structures of various anomalies observed in 
pressure-temperature or density-temperature phase diagrams
of liquid water and silica.~\cite{Errington:2001ha,
Shell:2002cw, Yan:2008un, Nayar:2013dl}
Recently, analogous nested structures were obtained using the
Stillinger--Weber potential for mW water, silicon, and germanium using 
Rosenfeld's scaling analysis.~\cite{Vasisht:2014jp, Dhabal:2016ha}
For core-softened potentials, the water-like nested structure of anomalies
have been intensely explored.~\cite{Errington:2006dl, Sharma:2006fy, deOliveira:2008ga,
BarrosdeOliveira:2010ib, Barraz:2011ht, Fomin:2011fz,
Fomin:2011gb, Salcedo:2011di, Fomin:2013ef, Fomin:2013fa}
However, these potential models exhibit positive slopes of the
coexistence lines between the HDL and LDL phases.
Thus, it is significant to explore the effect of the negative slope on
the cascade anomalies in the FJ model.
In this paper, we describe
the nested structures of observed anomalies in the FJ
model, which, to the best of our knowledge, had yet to be examined.
The hierarchical order of the anomalies is similar to that observed
in the water-like anomalies:
the excess entropy anomaly precedes dynamic anomalies,
which precede the density anomaly.~\cite{Errington:2001ha, Errington:2001ha}

This paper is organized as follows:
details of the molecular dynamics simulation using
the FJ model are explained in Sec. II;
numerical results of the simulation are presented and discussed in
Sec. III;
conclusions of the study are presented in Sec. IV.

\section{Model and simulation methods}

We carried out molecular dynamics simulations for a one-component
liquid system
using the FJ potential.~\cite{Abraham:2011bx, Gordon:2014fo, Sun:2015jj,
Sun:2017wo, Sun:2018ff}
Particles of mass $m$ interact via
\begin{widetext}
\begin{equation}
\phi(r)  = \epsilon\left[(a/r)^n +
 \frac{A_0}{1+\exp\left[\frac{A_1}{A_0}(r/a-A_2)\right]}
-\frac{B_0}{1+\exp\left[\frac{B_1}{B_0}(r/a-B_2)\right]}
\right],
\label{eq:fermi_jagla}
\end{equation}
\end{widetext}
with fixed parameters $n$, $A_i$, and $B_i$ $(i=0, 1, 2)$, whose
values are presented in Table~\ref{table:fermi_jagla}.
The parameters $a$ and $\epsilon$ represent the length and energy scales,
respectively, of the potential.

\begin{table}[t]
\small
  \caption{Parameters in Fermi--Jagla potential}
  \label{table:fermi_jagla}
  \begin{tabular*}{0.5\textwidth}{@{\extracolsep{\fill}}ccccccc}
    \hline
    $n$ & $A_0$ & $A_1$ & $A_2$ & $B_0$ & $B_1$ & $B_2$\\
    \hline
    20 & 4.56 & 28.88 & 1.36 & 1.00 & 3.57 & 2.36\\
    \hline
  \end{tabular*}
\end{table}

Our simulation system is composed of $N$ identical particles ($N=1728$) in a
cubic box of volume $V$ under periodic boundary conditions.
Throughout this paper, the numerical results are presented in units of 
$a$, $\epsilon/k_\mathrm{B}$, $\sqrt{ma^2/\epsilon}$ for length,
temperature, and time, respectively, where $k_\mathrm{B}$ is the
Boltzmann constant, and $T$ is the temperature.
Accordingly, the pressure $p$, diffusion constant $D$, and shear viscosity
$\eta$ are presented
in units of $\epsilon/a^3$, $a/\sqrt{m/\epsilon}$, and
$\sqrt{\epsilon/m}/a^2$, respectively.
A time step of $\Delta t=0.001$ and a cut-off length $r_c=4.0$ for the
potential were used in the simulations.
The investigated number densities and temperatures were $\rho=N/V\in[0.2,
1.0]$ and $T\in [0.11, 0.80]$, respectively.
We performed molecular dynamics simulations with an NVE ensemble for
$10^6$ steps at each
thermodynamic state to calculate
various thermodynamic and dynamic quantities after equilibrations
for $10^6$ steps in an NVT ensemble.
This time scale $t=10^3$ is sufficiently long to quantify transport
coefficients such as diffusion constant $D$ and shear viscosity $\eta$.

As performed in Ref.~\onlinecite{Abraham:2011bx}, we first calculated the
pressure-volume ($p$-$V$) curve for various states to
determine the phase diagram.
From the $p$-$V$ curve (data not shown), it can be seen that the thermodynamically unstable
states in the liquid phase 
developed with decreasing temperature, and the associated LLPT, which 
distinguishes the HDL and LDL phases, was determined.

To examine the dynamic properties, the diffusion constant $D$
was calculated from the Einstein relation,
\begin{equation}
D = \lim_{t\to\infty} \frac{\langle \Delta r(t)^2\rangle}{6t},
\end{equation}
where $\langle \Delta r(t)^2\rangle= (1/N)\langle \sum_{i=1}^N
(\bm{r}_i(t)-\bm{r}_i(0))^2\rangle$ is the mean square displacement.
Here, $\bm{r}_i(t)$ is the $i$-th particle position at time $t$.
In addition, the viscosity $\eta$ was calculated from the 
stress correlation function,
\begin{equation}
\eta_{\alpha\beta}(t) = \langle
 \sigma_{\alpha\beta}(t)\sigma_{\alpha\beta}(0)\rangle
\quad (\alpha, \beta = x, y, z)
\end{equation}
where $\sigma_{\alpha\beta}(t)$ is the stress tensor at time $t$.
From the Green--Kubo formula, the shear stress $\eta$ was quantified by
obtaining the averaged integral of the off-diagonal component,
\begin{equation}
\eta = \frac{1}{3k_\mathrm{B}TV} \int_0^\infty (\eta_{xy}(t) +
 \eta_{xz}(t)+\eta_{yz}(t)) dt.
\end{equation}
Note that we generated five independent trajectories to obtain the
convergence of the stress correlation function.
Furthermore, 
the long-time behavior of the averaged stress correlation function, $(\eta_{xy}(t) +
 \eta_{xz}(t)+\eta_{yz}(t))/3$, was fitted
by the stretched exponential function, $G_\mathrm{p}\exp[-(t/\tau_\eta)^{\beta}]$, with the plateau modulus
$G_\mathrm{p}$, the
stress relaxation time $\tau_\eta$, and the stretched exponent $\beta$.

Furthermore, the thermodynamic anomaly was characterized by the excess entropy, $S^\mathrm{e}$,
which is defined as the difference between the total
entropy, $S$, and its ideal gas component, $S^\mathrm{i}$.
The excess entropy was calculated following the procedure of the
thermodynamic integration described in Ref.~\onlinecite{Sciortino:2000fq}.
In practice, the calculation consists of two integrations.
Starting from the excess entropy at the
reference point $(V_\mathrm{ref},
T_\mathrm{ref})$, $S^{\mathrm{e}}_\mathrm{ref}$, the excess entropy at
the point $(V, T_\mathrm{ref})$ is given by
the integration along an isotherm:
\begin{equation}
S^\mathrm{e}(V, T_\mathrm{ref}) - S^\mathrm{e}_\mathrm{ref}= \frac{U(V,
 T_\mathrm{ref})}{T_\mathrm{ref}} + \int_{V_\mathrm{ref}}^{V}
 \frac{p^\mathrm{e}}{T_\mathrm{ref}} dV',
\end{equation}
where $p^\mathrm{e}$ and $U$ denote the excess pressure and
the potential energy, respectively.
Next, the entropy at any temperature is obtained by the integration
along the isochore:
\begin{equation}
S^\mathrm{e}(V, T) = S^\mathrm{e}(V, T_\mathrm{ref}) + \int_{T_\mathrm{ref}}^{T}
 \frac{C_V(T')}{T'} dT',
\end{equation}
where $C_V$ is the heat capacity.
In our calculations, the reference point was chosen as $({\rho_\mathrm{ref}}^{-1},
T_\mathrm{ref}) = (5.8, 2.0)$.
In addition, it is necessary to evaluate the excess entropy at the
reference point, $S^\mathrm{e}_\mathrm{ref}/Nk_\mathrm{B}=[
U_\mathrm{ref}/N + p^\mathrm{e}_\mathrm{ref} V_\mathrm{ref}/N -
\mu^\mathrm{e}_\mathrm{ref}]/T_\mathrm{ref}$.~\cite{Dhabal:2016ha}
Here, $\mu^\mathrm{e}_\mathrm{ref}$ denotes the excess chemical
potential at the reference point, which was determined by the Widom
particle insertion method.~\cite{Frenkel:2001tf}
The estimated values are $U_\mathrm{ref}/N\approx -2.648$, $p^\mathrm{e}_\mathrm{ref}\approx -0.091$,
$\mu^\mathrm{e}_\mathrm{ref}\approx -2.249$,
and  $S^\mathrm{e}_\mathrm{ref}/Nk_\mathrm{B}\approx -0.464$.

\section{Results}

\begin{figure}[t]
\centering
\includegraphics[width=0.35\textwidth]{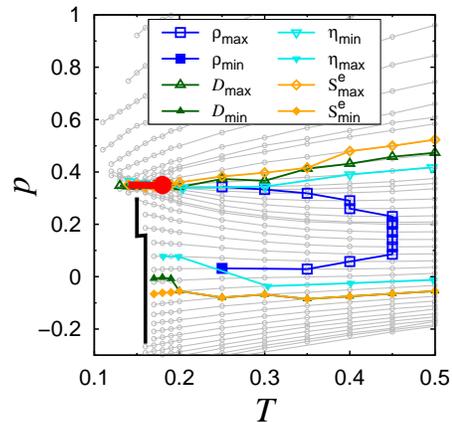}
\caption{Pressure-temperature phase diagram of the Fermi--Jagla model.
The water-like liquid-liquid transition critical point at $(p_c,
 T_c)\approx(0.35, 0.18)$ is represented by a red circle.
The black line represents the boundary temperatures, below which 
 crystallization spontaneously occurs.
From the critical point, a coexistence line with a weak negative slope, $dp/dT\approx -0.08$,
 between the high-density liquid and low-density liquid phases is represented by a red line.
The blue, green, cyan, and orange curves represent the loci of the maximum and minimum temperature
for the density $\rho$, diffusion $D$, shear viscosity $\eta$, excess entropy
 $S^\mathrm{e}$, respectively.
The open and closed symbols correspond to the maximum and minimum
 temperatures, respectively, at constant pressure.
}
\label{fig:phase_diagram}
\end{figure}

The phase diagram of the FJ model is shown in Fig.~\ref{fig:phase_diagram}.
This phase diagram reproduces the previous results of
Refs.~\onlinecite{Abraham:2011bx, Sun:2017wo} with the LLCP, $(p_c, T_c)\approx (0.35, 0.18)$, and the
coexistence line $dp/dT\approx -0.08$, between the HDL and LDL phases.
The weak but negative slope of the coexistence line implies that the
degree of ordering in the
HDL phase (high-pressure region) is lower than that in LDL phase
(low-pressure region).
This is due to the Clausius--Clapeyron equation, $dp/dT = \Delta
S/\Delta V$, where the volume difference, $\Delta
V=V_\mathrm{LDL}-V_\mathrm{HDL}>0$, leads to the entropy difference, $\Delta
S=S_\mathrm{LDL}-S_\mathrm{HDL}<0$, between the two phases.
Furthermore, this behavior is similar to that observed in simulations of
liquid water model, while the
opposite slope is obtained in the Jagla potential, which is a discontinuous
version of the FJ model.~\cite{Xu:2005ha}
This indicates that a small difference in the core-softened potential
causes such significant change in the slope $dp/dT$.
In fact, recent numerical simulations revealed that
the phase diagram near the LLCP is largely affected by
the depth and the distance of the potential minimum in the Jagla
model.~\cite{Gibson:2006cj, Luo:2014jw, Luo:2015ku, Ricci:2017ik}
Identifying the origin of the negative/positive slopes of LLPT coexistence lines in
core-softened potentials remains elusive.
The anomalous region regarding the density is also described by the locus of
the density maximum points in Fig.~\ref{fig:phase_diagram}.
The overall behavior is in accordance with that reported in the previous
studies.~\cite{Abraham:2011bx, Sun:2017wo}

\begin{figure*}[t]
\includegraphics[width=0.8\textwidth]{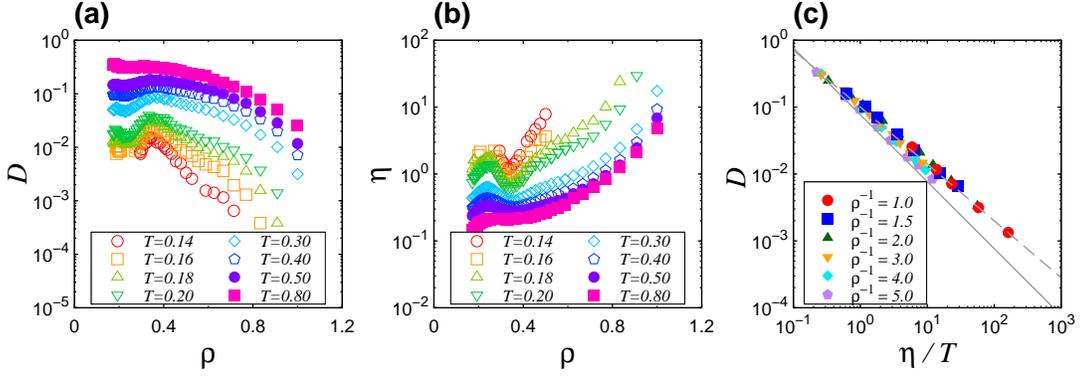}
\caption{Density $\rho$ dependence of (a) diffusion constant $D$, (b) shear
 viscosity $\eta$.
(c) Relationship between diffusion constant $D$ and shear viscosity
 $\eta$ divided by temperature $T$ for various number densities $\rho$.
The straight line satisfies the SE relation,
$D\sim (\eta/T)^{-1}$.
The dashed line represents the fractional SE formula, $D\sim
 (\eta/T)^{-\zeta}$ with the exponent $\zeta = 0.85$.}
\label{fig:dynamics_entropy}
\end{figure*}

\begin{figure*}[t]
\includegraphics[width=0.8\textwidth]{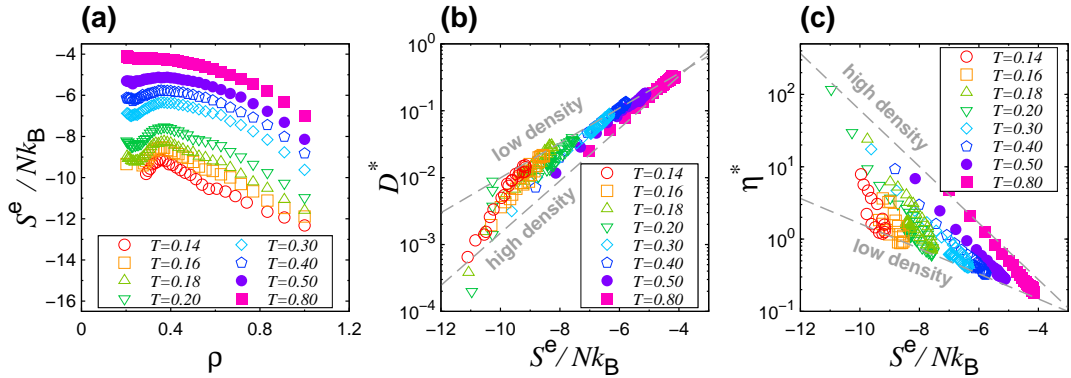}
\caption{
(a) Density $\rho$ dependence of  excess entropy $S^{\mathrm{e}}/Nk_\mathrm{B}$.
(b) Reduced diffusion constant, $D^*$,
and (c) shear viscosity, $\eta^*$, as a function of excess entropy,
 $S^{\mathrm{e}}/Nk_\mathrm{B}$.
The dashed lines are the Rosenfeld's scaling relationships: (a) $D^*=a_D\exp(b_D
 S^{\mathrm{e}}/Nk_\mathrm{B})$; (b) $\eta^*=a_\eta\exp(b_\eta
 S^{\mathrm{e}}/Nk_\mathrm{B})$.
The dashed lines are described with the slopes $(b_D, b_\eta)=(0.9,
 -0.9)$ (high density) and $(b_D, b_\eta)=(0.6,
 -0.4)$ (low density).
}
\label{fig:rosenfeld}
\end{figure*}

The anomalous properties in transport coefficients such as diffusion
constant $D$ and shear viscosity $\eta$ were investigated.
Figures~\ref{fig:dynamics_entropy}(a) and \ref{fig:dynamics_entropy}(b) show the density
dependence of $D$ and $\eta$, respectively, at various temperatures.
The anomalies are characterized by the regions $(\partial D/\partial
\rho)_T>0$ and $(\partial \eta/\partial \rho)_T<0$.
In Fig.~\ref{fig:phase_diagram}, the anomalous regions regarding $D$ and
$\eta$ are represented by the loci of the maximum and minimum points of
$D(\rho)$ and $\eta(\rho)$.
These anomalies appear at similar thermodynamic states; 
however, the region of the diffusion anomaly is slightly larger than
that of the viscosity anomaly.
In both cases, the transport anomalies 
surround the anomalous region of the density maximum, as shown in Fig.~\ref{fig:phase_diagram}.

These calculations also provide the information on the breakdown of 
the SE relationship, $D\sim \eta/T$, between the diffusion
constant $D$ and the shear viscosity $\eta$.
The breakdown of the SE relationship is evident in various glass-forming
liquids.~\cite{Sillescu:1999ci,
Ediger:2000ed, Shi:2013ji, Sengupta:2013dg}
This hydrodynamic anomaly is also demonstrated in simulations and
experiments on supercooled water.~\cite{Chen:2006kk, Kumar:2006hx, Becker:2006ju, Xu:2009hq,
Dehaoui:2015ii, Guillaud:2016bk, Galamba:2017eq, Kawasaki:2017gw}
For the Jagla potential model, 
the fractional SE relationship, $D\sim(\tau/T)^{-\zeta}$, with the
exponent $\zeta <1$ was demonstrated,
where the viscosity $\eta$ is replaced by the structural relaxation time
$\tau$.~\cite{Xu:2009hq}
The validity of such replacement becomes debatable when making an accurate
assessment of the SE relationship.~\cite{Shi:2013ji, Sengupta:2013dg}
Thus, the quantification of $\eta$ is necessary to characterize the SE
relationship using the core-softened potentials.
In Fig.~\ref{fig:dynamics_entropy}(c), 
the relationship $D\sim (\eta/T)^{-\zeta}$ in the FJ model is examined.
At $\rho^{-1}=5.0$, the temperature dependence exhibits a weak
violation of the SE relationship with $\zeta \ls 1$.
In contrast, the breakdown of the SE relationship becomes noticeable at higher
densities with $\zeta \approx 0.85$.
This change in the exponent $\zeta$ from LDL to HDL indicates
the connection between the violation of SE
relationship and the fragility, that is,
the degree of the Arrhenius behavior in the dynamic
property according to Angell's classification.~\cite{Angell:1995dp}
In fact, the temperature dependence of $D$ in HDL deviates from the Arrhenius
behavior with decreasing temperature, whereas that in LDL follows the
Arrhenius temperature dependence.~\cite{Abraham:2011bx, Sun:2017wo}
The connection between the fragility and the violation of the SE relationship
was also demonstrated in glass-forming liquids.~\cite{Ediger:2000ed,
Kawasaki:2013bg, Kawasaki:2014ky, Ozawa:2016bk}

\begin{figure}[t]
\centering
\includegraphics[width=0.35\textwidth]{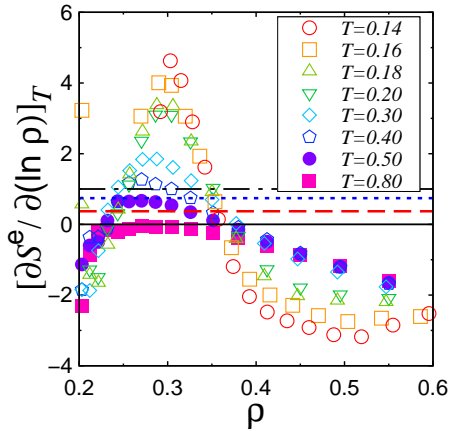}
\caption{Partial derivative of excess entropy, $S^{\mathrm{e}}$, with
 respect to density, $\ln \rho$, at constant temperature, $T$, against
 density, $\rho$.
The solid, dashed, dotted, and dot-dashed lines represent the onset values of structural,
 diffusion, viscosity, and density anomalies, respectively.
}
\label{fig:stability}
\end{figure}

Next, the thermodynamic anomaly was examined from the density dependence
of the excess entropy, $S^\mathrm{e}(=S-S^\mathrm{i})$,
which is shown in Fig.~\ref{fig:rosenfeld}(a).
As outlined in Refs.~\onlinecite{Errington:2006dl,
Nayar:2013dl, Vasisht:2014jp, Dhabal:2016ha},
the entropy anomaly is associated with the density anomaly from
the thermodynamic relation,
\begin{equation}
\left(\frac{\partial \rho}{\partial T}\right)_p =
 \rho^2\left(\frac{\partial \rho}{\partial
	p}\right)_T\left(\frac{\partial S}{\partial \rho}\right)_T, 
\end{equation}
and the thermodynamic stability condition, $(\partial \rho/\partial p)_T>0$.
Since the temperature and density dependence of the ideal gas entropy is
expressed as $S^\mathrm{i}/Nk_\mathrm{B}=-\ln
\rho+c(T)$ with the temperature dependent constant $c(T)$, the relation
\begin{equation}
\left[\frac{\partial S^\mathrm{e}}{\partial (\ln
 \rho)}\right]_T=\left[\frac{\partial S}{\partial (\ln \rho)}\right]_T+1,
\end{equation}
can be obtained.
Thus, the density anomaly is characterized by 
\begin{equation}
\left[\frac{\partial S^\mathrm{e}}{\partial (\ln\rho)}\right]_T>1.
\label{eq:density_anomaly}
\end{equation}
In Fig.~\ref{fig:phase_diagram}, the anomalous region is described by
the loci of the maximum and minimum of $S^{\mathrm{e}}(\rho)$.
It has been demonstrated that 
the anomalous region of the excess entropy is observed at the outermost
boundary over the density and transport anomalies.
Thus, the excess entropy anomaly is associated with the
structural order anomaly caused by the LLPT between the HDL and LDL phases.
In fact, the excess entropy, $S^\mathrm{e}$, is conventionally approximated 
as the two-body excess entropy $S_2$ calculated from the integral of the radial distribution function $g(r)$.

To correlate the observed density, dynamical, and entropy anomalies, 
the scaling relationship between the transport coefficients and the excess
entropy in the FJ model was examined.
More specifically, Rosenfeld's scaling was utilized in an exponential form,
\begin{equation}
X^* = a_X \exp(b_X S^\mathrm{e}),
\label{eq:rosenfeld}
\end{equation}
between an arbitrary dimensionless transport coefficient $X^*$ and the
excess entropy $S^\mathrm{e}$.~\cite{Rosenfeld:1977tu, Rosenfeld:1999ii}
In Eq.~(\ref{eq:rosenfeld}), $a_X$ and $b_X$ are the parameters.
The reduced diffusion constant and shear
viscosity can be expressed as
$D^*=D\rho^{1/3}/(k_\mathrm{B}T/m)^{1/2}$ and $\eta^*=\eta
\rho^{-2/3}/(mk_\mathrm{B}T)^{1/2}$, respectively.
Figures~\ref{fig:rosenfeld}(b) and \ref{fig:rosenfeld}(c) shows the semi-log plots of $D^*$ and $\eta^*$ against
$S^\mathrm{e}$ for different isotherms.
In addition, Rosenfeld's scaling relationship in
Eq.~(\ref{eq:rosenfeld}) is examined.
Instead of a single exponential scaling, 
two separate branches are observed.
One corresponding to high density (non-anomalous) region and the other corresponds to
low density (anomalous) region.
Similar two-branched scaling of Rosenfeld's relationship
has been demonstrated in recent studies using the 
Stillinger--Weber potential.~\cite{Vasisht:2014jp, Dhabal:2016ha}
Note that the absolute values of the coefficients $b_D$ and $b_\eta$ become larger with
increasing the density.
This tendency is seemingly contrary to results reported in
Refs.~\onlinecite{Vasisht:2014jp, Dhabal:2016ha}.

The slope obtained from the plot of Rosenfeld's scaling relationship
enables us to examine the
relationship between the excess entropy anomaly and the
transport anomalies in the FJ model.
As reported in Ref.~\onlinecite{Errington:2006dl}, the region of the transport anomaly
is represented by the partial derivative of the excess entropy
$S^\mathrm{e}$ with respect to the logarithmic density $\ln \rho$,
\begin{equation}
\left[\frac{\partial S^\mathrm{e}}{\partial (\ln \rho)}\right]_T > c,
\end{equation}
where the constant $c$ is given by the slope of Rosenfeld's scaling
relationship.
In practice, the conditions $c=1/3b_D$ and $c=2/3|b_\eta|$ correspond to the diffusion
and viscosity anomalies, respectively.
On the other hand, $c=0$ represents the criterion of the excess entropy anomaly.
Furthermore, $c=1$ corresponds to the condition for the onset of density anomaly, 
as indicated in Eq.~(\ref{eq:density_anomaly}).
In~Fig.~\ref{fig:stability}, the partial derivative $[\partial S^\mathrm{e}/\partial (\ln
\rho)]_T$ is plotted as a function of the density $\rho$ at various temperatures.
The values of $c$ for the criteria of the excess entropy, diffusion,
viscosity, and density anomalies are also represented by the horizontal lines.
Note that the high density values of the slope of the Rosenfeld's
scaling are used for both the diffusion and the viscosity (see
Figs.~\ref{fig:rosenfeld}(b) and ~\ref{fig:rosenfeld}(c)).
This analysis allows us to estimate the onset of the anomalies in the $p$-$T$
phase diagram from non-anomalous branches.~\cite{Vasisht:2014jp, Dhabal:2016ha}
In fact, the order of the anomalies in Fig.~\ref{fig:phase_diagram} from
the outermost boundary to the innermost boundary is consistent with that demonstrated
in Fig.~\ref{fig:stability}, that is, the anomaly in the excess
entropy is followed by the diffusion anomaly, then
the viscosity anomaly, and, finally, density anomaly.
The hierarchy of the water-like anomalies of the FJ model
is thus unveiled by the demonstrated nested dome-like structures.

\section{Conclusions}

We performed molecular dynamics simulations using the FJ model,
which is one of the core-softened potentials with two length scales.
Our results show a water-like LLPT between the HDL and LDL phases in the
$p$-$T$ phase diagram, which is consistent with the results of previous
studies,~\cite{Abraham:2011bx, Sun:2017wo}
In particular, the slope of the LLPT line is weakly negative, which is
an important characteristic of the liquid-water anomalies.
We also demonstrated that the density anomaly boundary emerges from the
LLCP.

We numerically calculated the transport coefficient such as
self-diffusion and shear viscosity.
From these calculations, the dynamic anomalies in the FJ model were
thoroughly characterized.
In addition, the violation of the SE relationship was clarified in the FJ
model.
The connection between the fragility and the degree of SE breakdown was
demonstrated, which is analogous to that exhibited in glass-forming liquids.

Furthermore, we characterized the thermodynamic anomalies from the excess
entropy, which was obtained by the thermodynamic integration.
From these calculations, the dynamic and thermodynamic anomalies were
thoroughly characterized as the dome-like boundaries in the phase diagram.
The nested structure of the anomaly boundaries implies the
connection between the thermodynamic and dynamic anomalies arising from the LLCP.

To unveil the connection, we utilized the Rosenfeld's scaling relationship
between the excess entropy and the dimensionless transport coefficients.
The slope of the exponential representation of the scaling is
related to the criteria for the appearance of the anomalies in the phase
diagram. 
The elucidated results are in accordance with the density dependence of the partial
derivative of the excess entropy with respect to the logarithmic
density along isotherms.
The anomalous regions of the FJ model, showing the negative
slope of the coexistence line between HDL and LDL,
follow the hierarchy of density, then diffusion,
viscosity, and, finally, structure with increasing the temperature.
This order is also compatible with that observed in the liquid-water model and
other water-like liquids.\cite{Errington:2001ha, Errington:2006dl}

It is worth mentioning that analogous nested structures have been
reported using
Stillinger--Weber potential for the models of mW water, silicon, and
germanium.~\cite{Vasisht:2014jp, Dhabal:2016ha}
These studies revealed that
the hierarchy is inverted between diffusion and excess
entropy, that is, the anomalous region of diffusion covers those of density and excess
entropy.
This sequence is consistent with that observed in silica.~\cite{Shell:2002cw}
In addition, the silica-like hierarchy has been obtained by increasing
the potential depth of the attraction part in
other core-softened potential.~\cite{Fomin:2013ef}
Although the form of this potential resembles the FJ model, 
the similarity and the differences is still elusive.
Further investigations extending to the generalization of the FJ model are currently underway.

\begin{acknowledgments}
The authors thank M. Ozawa for helpful discussions.
This work was supported by JSPS KAKENHI Grant Number JP16H00829 on
Innovative Areas (2503) Studying the Function of Soft Molecular Systems
by the Concerted Use of Theory and Experiment.
The numerical calculations were performed at Research Center of Computational
Science, Okazaki Research Facilities, National Institutes of Natural Sciences, Japan.
\end{acknowledgments}

\end{document}